\documentclass[final]{aipproc} 
\layoutstyle{8x11double} 
\usepackage{amssymb} 
\usepackage{graphicx} 
\usepackage{graphics} 
\usepackage{epsfig} 
 
\begin{document} 
\title{Application of the SALI chaos detection method to accelerator mappings} 
 
\author{T.~Bountis$^a$ and Ch.~Skokos$^{a,}$$^{b,}$$^{c,}$\footnote{E-mails: 
 bountis@math.upatras.gr (T.B.), hskokos@cc.uoa.gr (Ch.S.)}} 
 {address={$^a$ Department of Mathematics, Division of Applied 
Analysis and Center for Research and Applications of Nonlinear 
Systems (CRANS), University of Patras, GR-26500 Patras, Greece 
\\ 
$^b$ Research Center for Astronomy and Applied Mathematics, 
Academy of Athens, Soranou Efesiou 4, GR-11527, Athens, Greece 
\\ 
$^c$ Department of Applications of Informatics and Management in 
Finance, Technological Institute of Mesologhi, GR--$30200$, 
Mesologhi, Greece}}

\begin{abstract} 
We apply the Smaller ALignment Index (SALI) method to a 
4--dimensional mapping of accelerator dynamics in order to 
distinguish rapidly, reliably and accurately between ordered and 
chaotic motion. The main advantage of this index is that it tends 
{\it exponentially} to zero in the case of chaotic orbits, while 
it fluctuates around non--zero values in the case of quasiperiodic 
trajectories. Thus, it avoids the notorious ambiguities concerning 
the eventual convergence of (maximum) Lyapunov exponents to 
(positive) non-zero values. Exploiting the different behavior of 
SALI in these two cases we produce phase space `charts' where 
regions of chaos and order are clearly identified. Evaluating the 
percentage of chaotic and escaping orbits as a function of the 
distance from the origin we are able to estimate rapidly and 
accurately the boundaries of the {\it dynamical aperture} of a 
proton beam, passing repeatedly through an array of magnetic 
focusing elements. 
\end{abstract} 
\maketitle 
\section {Introduction} 
\label{INTRO} 
 
One of the basic problems in accelerator physics is the 
determination of the stability regions (islands of quasiperiodic 
motion) of a particle beam, as well as its {\it dynamical 
aperture}, i.e the domain about the ideal circular path in which 
the particles' motion remains bounded \cite{GST97}. In the case of 
`flat' hadron beams, where the horizontal (x-) motion is dominant, 
2--dimensional (2D) area--preserving mappings are frequently used 
to model the effect of nonlinearities as the particles repeatedly 
pass through focusing magnetic elements \cite{BTTS94}, or 
experience the beam-beam interaction with particles of a colliding 
beam \cite{BBE84}. The presence of invariant curves around the 
origin of such mappings (corresponding to the particle's ideal 
circular orbit), guarantees the long--time stability of the beam. 
In that case, the chaotic motion that exists between these 
invariant curves remains always bounded and so the beam particles 
do not escape to infinity. On the other hand, in the case of 
`elliptical' bunched hadron beams, where the vertical (y-) and 
longitudinal (z-) motion cannot be neglected and which are 
modelled by $k=2M$--dimensional symplectic mappings with $M>1$, 
chaotic regions can be connected, providing a path for the orbits 
to move away from the origin and eventually escape to infinity. 
This practically means the loss of particles in the storage rings 
of the accelerator and hence a reduction of the beam's {\it 
dynamical aperture}. 
 
In the present paper, we face the problem of the estimation of the 
dynamical aperture of a 4D symplectic mapping, which describes the 
motion of a hadron particle as it passes through a magnetic 
focusing element of the FODO cell type. This system has already 
been studied in \cite{BT91,BK94,VBK96,VIB97}. In particular 
Vrahatis et al. \cite{VIB97} tried to approximate invariant tori 
existing as far away from the origin as possible by computing 
stable periodic orbits of very high period. Their idea was that 
the corresponding islands of stability and the quasiperiodic 
`tori' around these periodic orbits could play the role of an 
effective barrier of orbital diffusion, although their presence 
does not exclude the possible `leaking' of chaotic orbits to large 
distances from the origin. 
 
A more direct approach to the problem of estimating the size of 
the dynamical aperture, is the actual characterization of orbits 
on a fine grid around the origin as ordered or chaotic (escaping 
or not escaping to infinity). In this way one can determine the 
region about the ideal circular path where predominantly ordered 
orbits exist, thus guaranteeing the stability of the beam, at 
least up to the number of iterations that the orbits have been 
computed. 
 
This approach requires the use of a fast and reliable method which 
can distinguish between ordered and chaotic motion rapidly, 
reliably and accurately. The usual method of the computation  of 
the maximal Lyapunov exponent \cite{BGGS80a,BGGS80b} does not meet 
these requirements as the number of iterations needed for the 
convergence of the Lyapunov exponent to its actual (zero or 
positive) value is not known a--priori and may become very high. 
Therefore, the application of this procedure to a huge number of 
initial conditions becomes impractical and its efficiency as a 
tool for studying the dynamical aperture of particle beams seems 
seriously doubtful . For these reasons, we prefer apply to our 
system the Smaller ALignment Index (SALI) method 
\cite{S01,SABV03a,SABV03b,SABV04}, which has been proved to be an 
efficient, reliable and very fast method of chaos detection.

\section {The SALI method} 
\label{SALI} 
 
The SALI method was introduced in \cite{S01} and has already been 
applied successfully to distinguish between ordered and chaotic 
motion in various mappings and Hamiltonian systems 
\cite{SABV03a,SABV03b,SABV04}, as well as problems of Celestial 
Mechanics \cite{S03,SESS04}, Galactic Dynamics \cite{MA05a,MA05b}, 
Field Theory \cite{AHHN05}  as well as non--linear lattices 
\cite{PBS04,ABS05}. 
 
In order to recall the definition of the SALI let us consider the 
$k$-dimensional phase space of a conservative dynamical system, 
e.~g.~a $2M$--dimensional symplectic mapping or a Hamiltonian flow 
of $N$ degrees of freedom, with $k=2N$. In a symplectic mapping 
the evolution of an orbit with initial condition $X(0)=(x_1(0), 
x_2(0), \ldots x_k(0))$, $k=2M$ is governed by the discrete--time 
equations of the mapping, having the form 
\begin{equation} 
X(n+1)=F(X(n)), 
\label{eq:disc_time} 
\end{equation} 
where $X(n)=(x_1(n), x_2(n), \ldots, x_k(n))$ is the orbit's 
location at the $n$--th iteration of the mapping. On the other 
hand, in a Hamiltonian flow, the motion of an orbit with initial 
condition $X(0)=(x_1(0), x_2(0), \ldots x_k(0))$, $k=2N$ is 
governed by Hamilton's equations of motion, which have the general 
form 
\begin{equation} 
\frac{d X(t)}{dt}=F(X(t)), 
\label{eq:cont_time} 
\end{equation} 
where $X(t)=(x_1(t), x_2(t), \ldots, x_k(t))$ is the orbit's position in the 
phase space at time $t$. 
 
Suppose we wish to determine the chaotic vs. ordered nature of an 
orbit $X(n)$ of a symplectic mapping (or $X(t)$ of a Hamiltonian 
system) with initial condition $X(0)$. To do so, one traditionally 
follows the evolution of one deviation vector $V(n)=(dx_1(n), 
dx_2(n), \ldots, dx_k(n))$ (or $V(t)$), which can be considered as 
initially pointing to an orbit nearby the one under study, and 
computes the orbit's maximal Lyapunov exponent 
\cite{BGGS80a,BGGS80b}. In the case of mappings the evolution of 
such a deviation vector $V(n)$ is governed by the equations of 
motion of the so--called tangent map: 
\begin{equation} 
V(n+1)=DF(X(n))\cdot V(n), 
\label{eq:tang_map} 
\end{equation} 
while, in the case of Hamiltonian flows we use the set of linear ordinary 
differential equations called variational equations: 
\begin{equation} 
\frac{dV(t)}{dt}=DF(X(t))\cdot V(t), 
\label{eq:var_eq} 
\end{equation} 
where $DF$ denotes the Jacobian matrix of equations 
(\ref{eq:disc_time}) or (\ref{eq:cont_time}) evaluated at the 
points of the orbit under study. Since, in the present paper, we 
study symplectic mappings, our notation from now on will be 
restricted to difference equations, although the following 
concepts can be easily extended to the case of differential 
equations describing Hamiltonian flows. 
 
For the evaluation of the SALI we follow the time evolution of two initially 
different deviation vectors $V_1(0)$, $V_2(0)$ and define SALI \cite{S01} as: 
\begin{equation} 
\mbox{SALI}(n)=\min \{ \|\hat{V}_1(n)+ \hat{V}_2(n)  \|, \|\hat{V}_1(n)- 
\hat{V}_2(n)  \|   \} 
\label{eq:SALI} 
\end{equation} 
where $\|\cdot \|$ denotes the usual Euclidean norm and 
$\hat{V}_i$, $i=1,2$ are normalized vectors with norm equal to 1, 
i.~e.~$\hat{V}_i(n)=V_i(n)/\| V_i(n)\|$. 
 
Two different behaviors of SALI are thus distinguished: 
\begin{enumerate} 
\item If the orbit under study is chaotic, the two vectors $\hat{V}_1(n)$, $\hat{V}_2(n)$ tend to coincide (or 
  become opposite) along the most unstable direction (corresponding to the maximal Lyapunov exponent). 
  In this case, SALI tends exponentially to zero following a rate which depends on the difference 
  between the two largest Lyapunov exponents \cite{SABV04}. 
\item If the orbit is ordered (quasiperiodic), there is no unstable direction and vectors  $\hat{V}_1(n)$, 
  $\hat{V}_2(n)$ tend to become tangent to the corresponding torus, having in general 
  different directions. In this case SALI remains different from zero, fluctuating around some mean value 
  \cite{SABV03b}. An exception to this behavior appears for ordered orbits of 
  2D mappings where the SALI tends to zero following a power law (see \cite{S01} for more details). 
\end{enumerate} 
 
The simplicity of SALI's definition, its completely different 
behavior for ordered and chaotic orbits and its rapid convergence 
to zero in the case of chaotic motion are the main advantages that 
make SALI an ideal chaos detection tool, perfectly suited for 
multidimensional conservative systems, such as proton (antiproton) 
beams in accelerator storage rings.

\section {Global dynamics of a 4D accelerator mapping} 
\label{MAP} 
 
Consider the 4D symplectic mapping: 
\begin{eqnarray} 
\left( \begin{array}{c} 
x_1(n+1) \\x_2(n+1) \\x_3(n+1) \\x_4(n+1) 
\end{array} \right) & = & 
\left( \begin{array}{cccc} 
\cos \omega_1 & -\sin \omega_1 & 0 & 0 \\ 
\sin \omega_1 & \cos \omega_1 & 0 & 0 \\ 
0 & 0 & \cos \omega_2 & -\sin \omega_2 \\ 
0 & 0 & \sin \omega_2 & \cos \omega_2 
\end{array} \right) \nonumber \\ 
 & \times & \left( \begin{array}{c} 
x_1(n) \\ 
x_2(n)+x_1^2(n)-x_3^2(n) \\ 
x_3(n) \\ 
x_4(n)-2 x_1(n) x_3(n) 
\end{array} \right), 
\label{eq:map} 
\end{eqnarray} 
describing the instantaneous sextupole nonlinearities experienced 
by the dynamics of a proton beam as it passes repeatedly through 
magnetic focusing elements of the FODO cell type 
\cite{BT91,BK94,VBK96,VIB97}. Here $x_1$ and $x_3$ are the 
particle's horizontal and vertical deflections from the ideal 
circular orbit of the beam, $x_2$ and $x_4$ are the associated 
momenta and $\omega_1$ and $\omega_2$ are related to the 
accelerator's tunes $q_x$, $q_y$ by 
\begin{equation} 
\omega_1 = 2 \pi q_x\, \, , \omega_2 = 2 \pi q_y. 
\label{eq:tunes} 
\end{equation} 
 
\begin{figure} 
\centerline{%
\begin{tabular}{c} 
\includegraphics[width=6.cm]{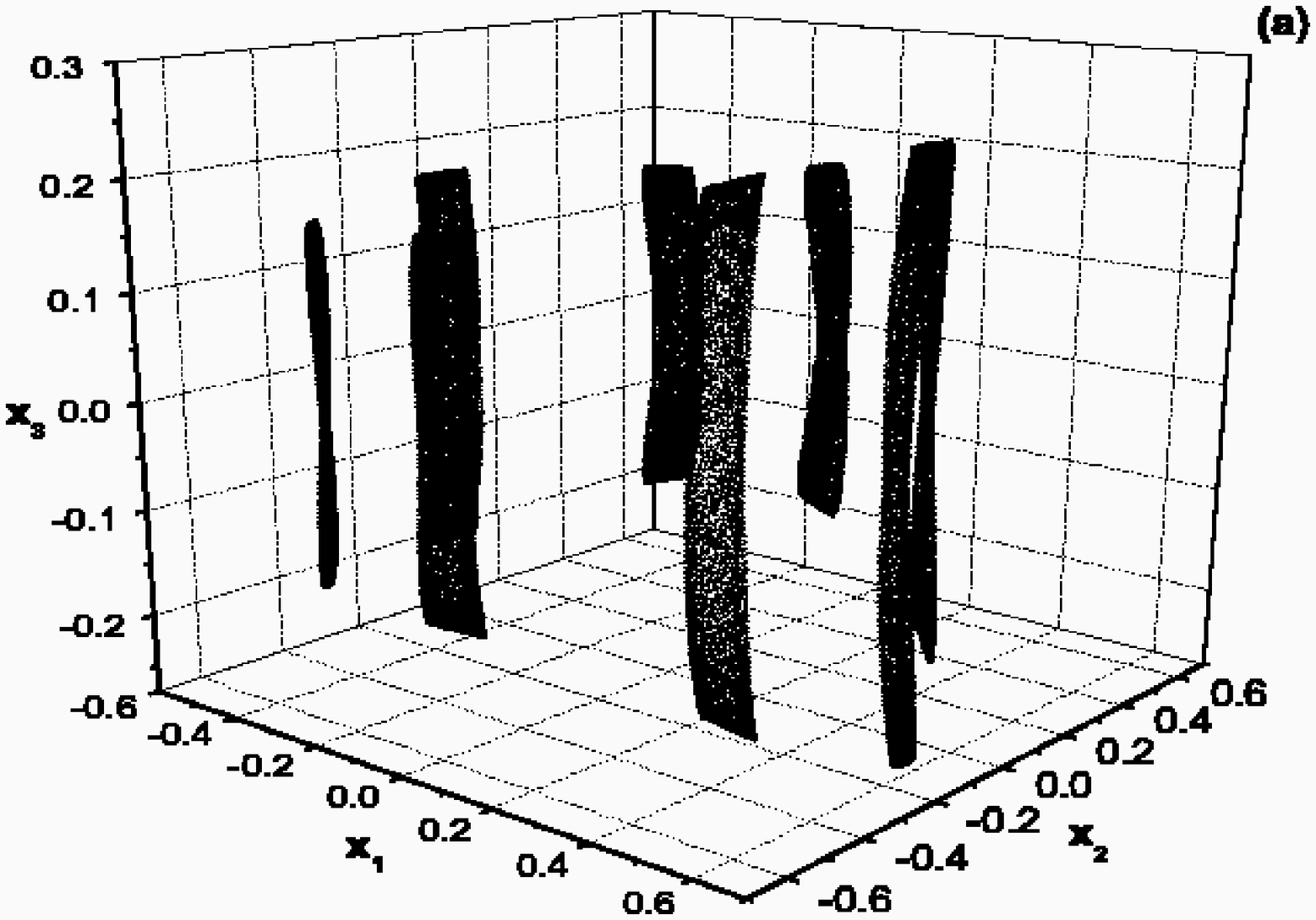} \\ 
\includegraphics[width=8cm]{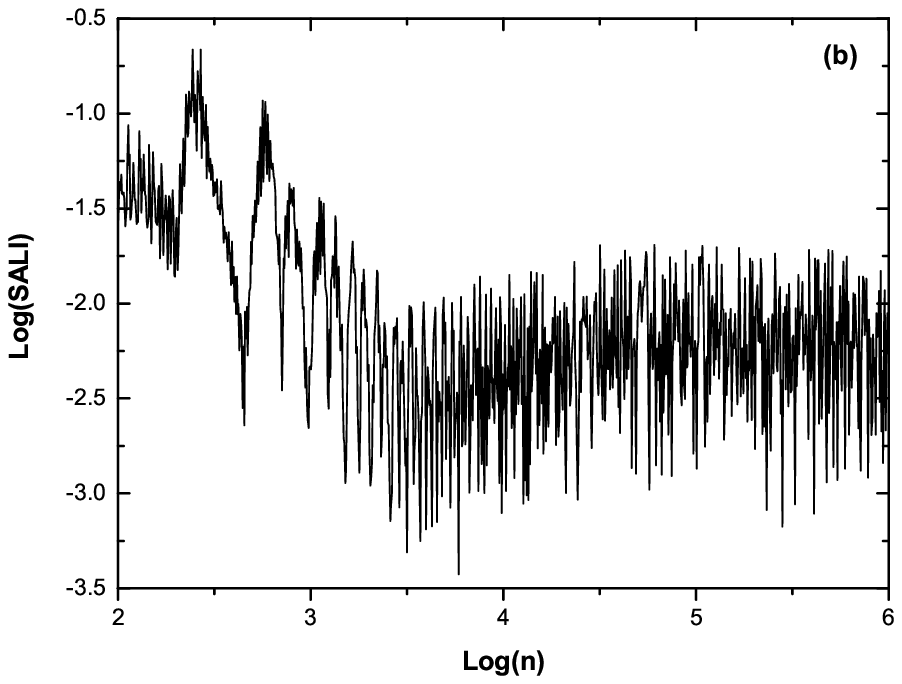} \\ 
\includegraphics[width=8cm]{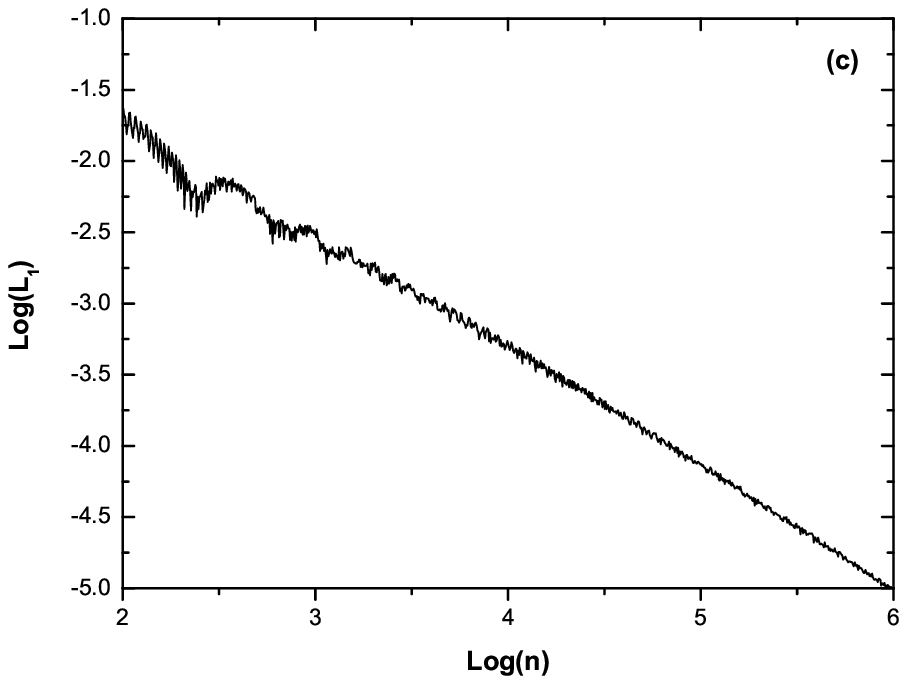} 
\end{tabular}} 
\caption{An ordered orbit of mapping (\ref{eq:map}). 
  (a) Projection of 50000  consequents of the orbit into the $x_1-x_2-x_3$ 
  space. Evolution of orbit's SALI 
  (b) and $L_1$ (c) as a function of mapping's iterations $n$ in log--log scale.} 
\label{fig:oo} 
\end{figure}

Let us first examine the behavior of the SALI for some individual 
orbits. Vrahatis et al. \cite{VIB97} have computed near the 
boundary of escape of the mapping several stable periodic orbits 
of very long period, as well as some invariant tori near them. In 
Figure \ref{fig:oo}(a) 
we see the projection into the $x_1-x_2-x_3$ space of one such 
ordered orbit of the mapping (\ref{eq:map}) for $q_x=0.61803$ and 
$q_y=0.4152$ first presented in figure 1(b) of \cite{VIB97}. This 
orbit is generated by a small perturbation in $q_x$ of the stable 
periodic orbit of period 13237 found in \cite{VIB97} for 
$q_x=0.61903$ and it lies on 8 tori in the $x_1-x_2-x_3$ space. 
The exact values of the orbit's initial condition, which will be 
denoted from now on as $X^*(0)=(x_1^*(0),x_2^*(0), x_3^*(0), 
x_4^*(0))$, can be found in table 3 of \cite{VIB97}.  Following 
\cite{BGGS80a,BGGS80b}, we may compute the maximal Lyapunov 
exponent $\lambda_1$, of the orbit as the limit for $n\rightarrow 
\infty$ of the quantity 
\begin{equation} 
L_1(n) = \frac{1}{n} \ln \frac{\|V(n) \|}{\|V(0) \|}, \,\, 
\mbox{i.~e.~} \, \, \lambda_1= \lim_{n\rightarrow \infty} L_1(n), 
\label{eq:Lyap} 
\end{equation} 
where $V(0)$, $V(n)$ are deviation vectors from the orbit at $n=0$ 
and $n>0$ iterations respectively. We recall that $\lambda_1=0$ 
for ordered orbits, while $\lambda_1 > 0$ for chaotic orbits. The 
ordered nature of the orbit is clearly revealed by the evolution 
of its SALI (Figure \ref{fig:oo}(b)) and of its $L_1$ (Figure 
\ref{fig:oo}(c)). The SALI remains different from zero fluctuating 
around $10^{-2.5}$ converging to this value long before $L_1$ 
becomes zero. Thus, as far as SALI is concerned, the computation 
could have been stopped after about 10000 iterations, concluding 
correctly that the orbit is ordered. 
 
\begin{figure} 
\centerline{%
\begin{tabular}{c} 
\vspace{0cm}  \includegraphics[width=7cm]{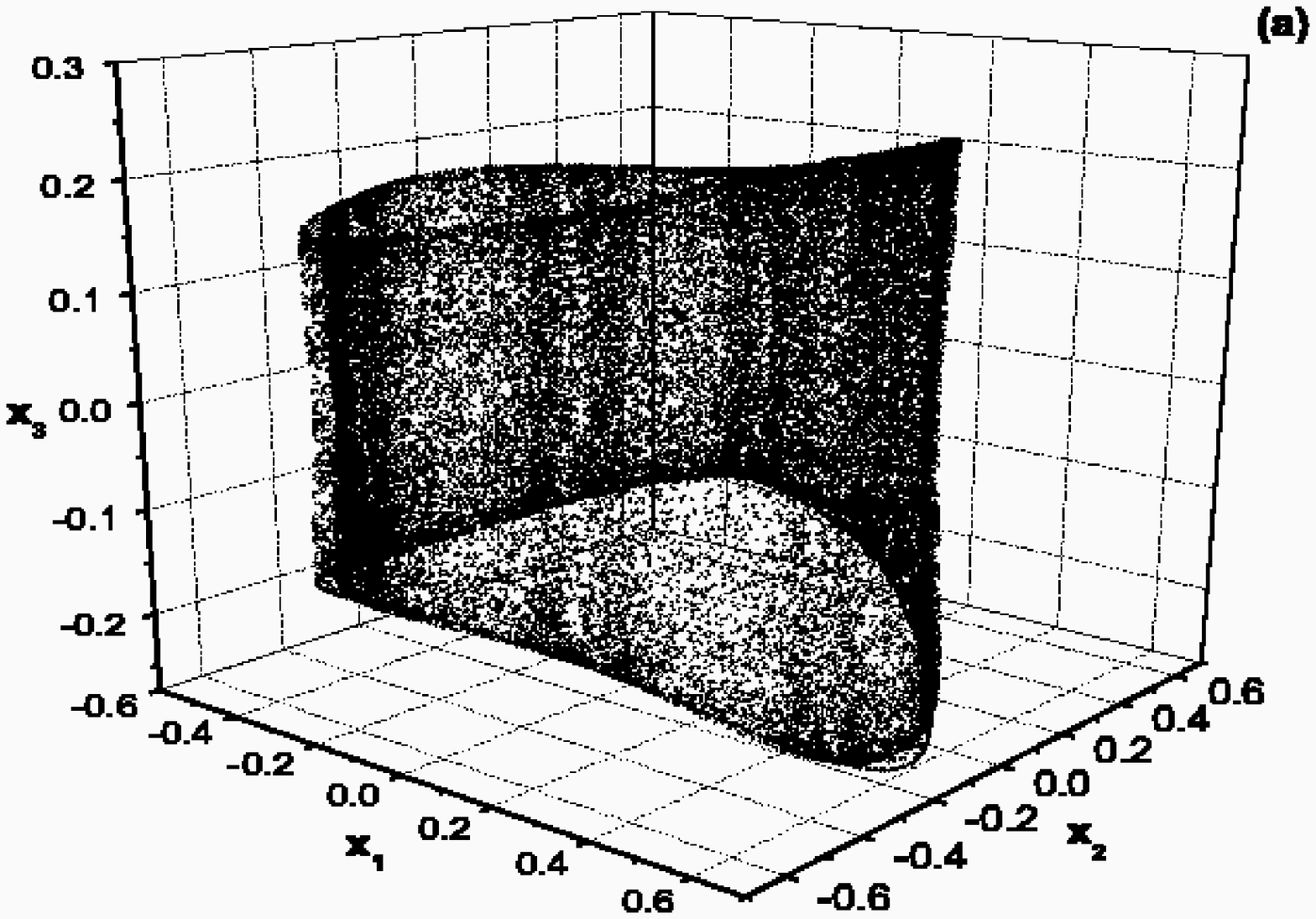} \\ 
\vspace{0cm}  \includegraphics[width=7.3cm]{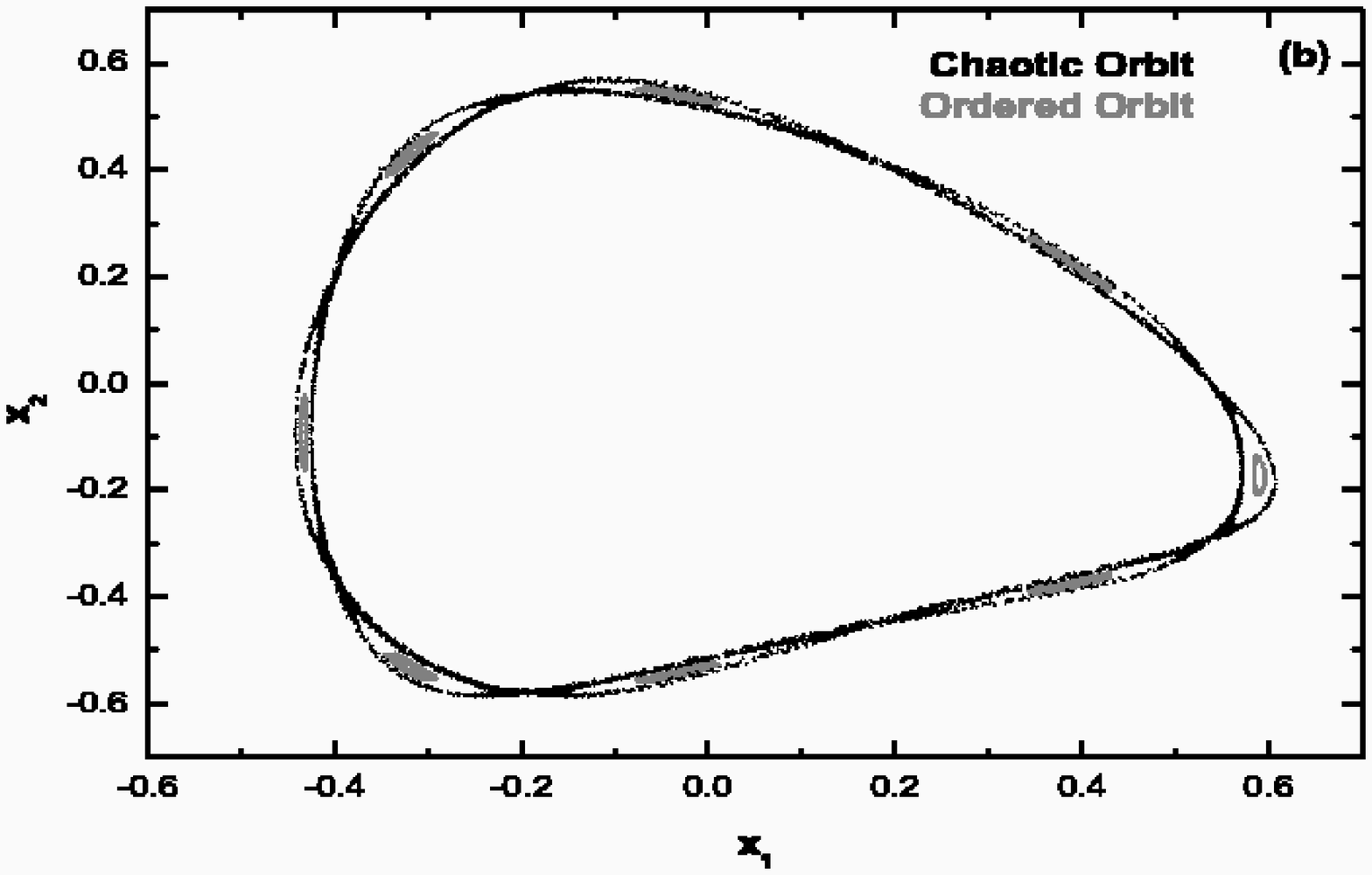} \\ 
\vspace{-1cm}  \includegraphics[width=7.5cm]{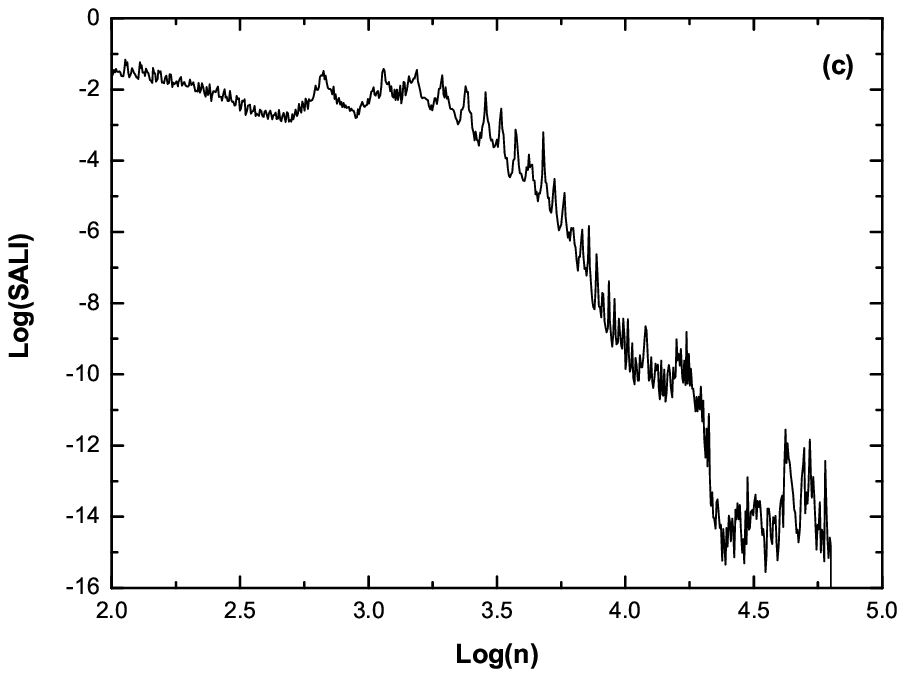} \\ 
\vspace{0.5cm} \includegraphics[width=7.5cm]{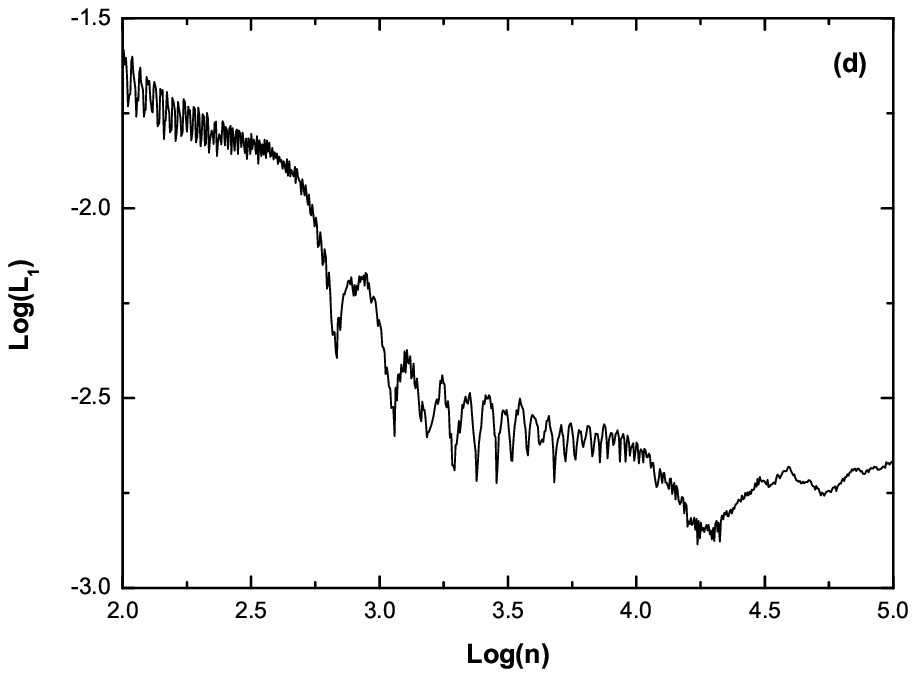} 
\end{tabular}} 
\caption{A chaotic orbit of mapping (\ref{eq:map}). 
  (a) Projection of 50000  consequents of the orbit into the $x_1-x_2-x_3$ 
  space. (b) Projection on the $x_1-x_2$ plane of the points of panel (a) 
  (black points) and 
  of Figure \ref{fig:oo}(a) (gray points) with 
  $|x_3|\leq 0.04$. Evolution of orbit's SALI 
  (c) and $L_1$ (d) as a function of mapping's iterations $n$ in log--log scale.} 
\label{fig:co} 
\end{figure}

By changing the $x_1$ coordinate of the initial condition of the 
orbit of Figure \ref{fig:oo}, to $x_1=0.624$ we get the weakly 
chaotic orbit plotted in Figure \ref{fig:co}(a), 
surrounding the 8 tori of Figure \ref{fig:oo}(a). This behavior is 
clearly seen in Figure  \ref{fig:co}(b) where we project on the 
$x_1-x_2$ plane the points of Figures  \ref{fig:oo}(a) and 
\ref{fig:co}(a) with $|x_3|\leq 0.04$. The SALI of the chaotic 
orbit decreases rapidly to zero, after a transient initial phase 
(Figure \ref{fig:co}(c)) reaching the limits of the computer's 
accuracy (i.~e.~$10^{-16}$) after about $n=20000$ iterations, 
showing clearly the chaotic nature of the orbit. We can of course 
set a less demanding threshold for the SALI's values in order to 
define an orbit as chaotic. Considering for example as such a 
threshold the value $\mbox{SALI}\approx 10^{-8}$, we can 
characterize the orbit as chaotic after only 8000 iterations. For 
the same number of iterations, $L_1$ (Figure \ref{fig:co}(d)) does 
not seem to converge to a non--zero value, so that many more 
iterations are needed for the definite characterization of the 
orbit as chaotic by the use of the maximal Lyapunov exponent. In 
fact, at about $10000$ iterations the maximal Lyapunov exponent 
gives an erroneous picture, as it starts to fall to values closer 
to zero! 
 
\begin{figure} 
\centerline{%
\begin{tabular}{c} 
\includegraphics[width=7cm]{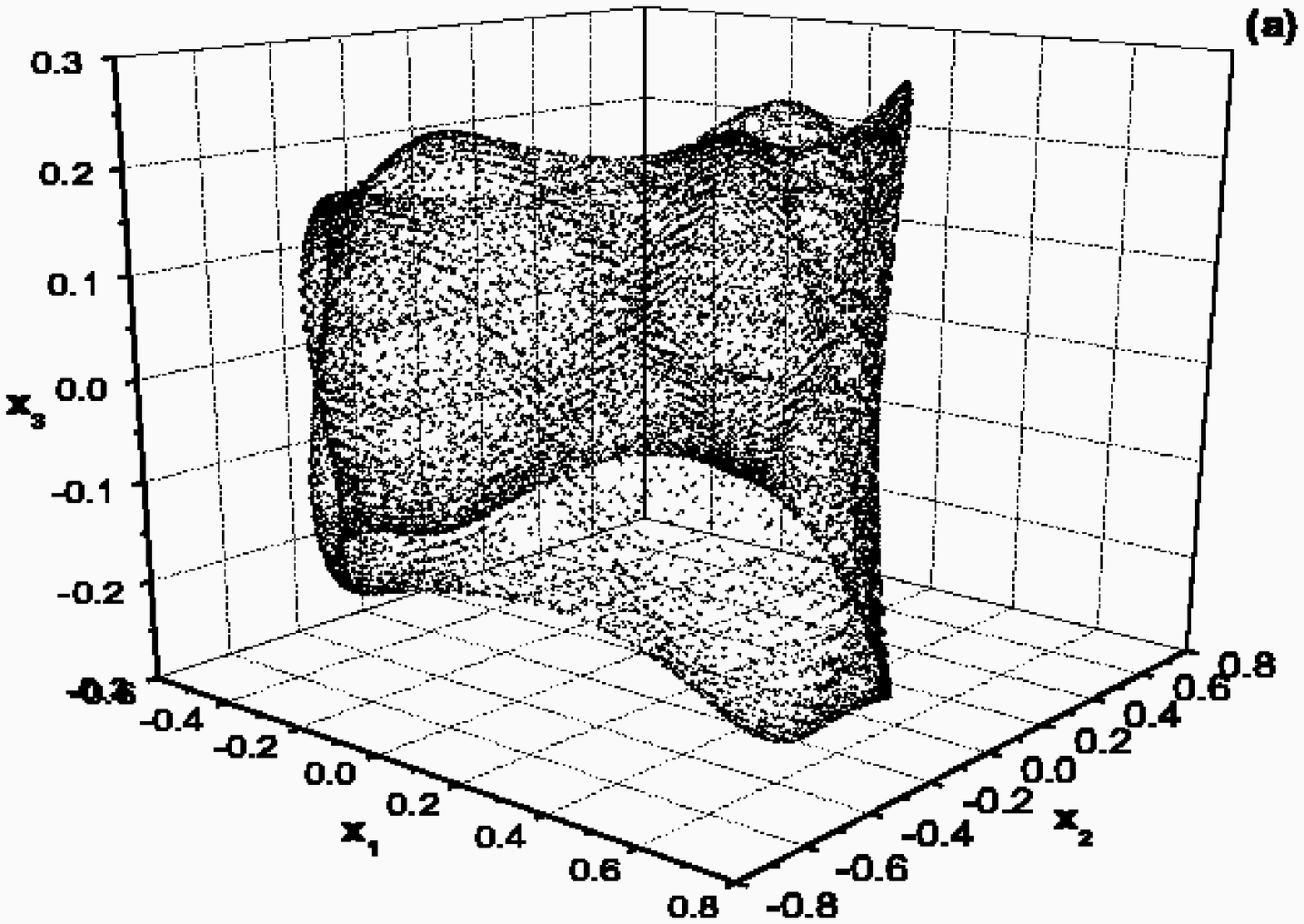} \\ 
\includegraphics[width=8cm]{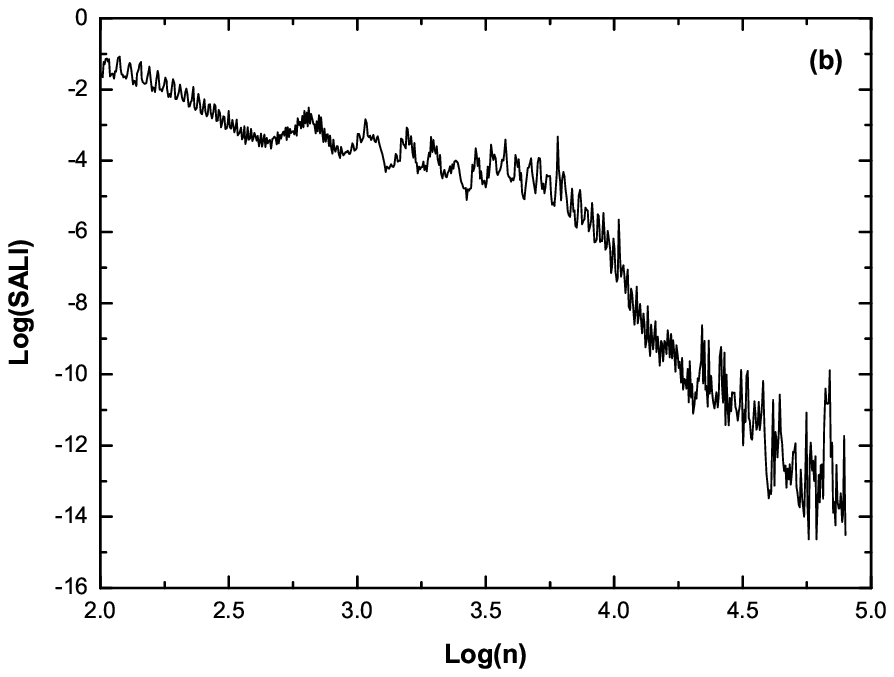} \\ 
\includegraphics[width=8cm]{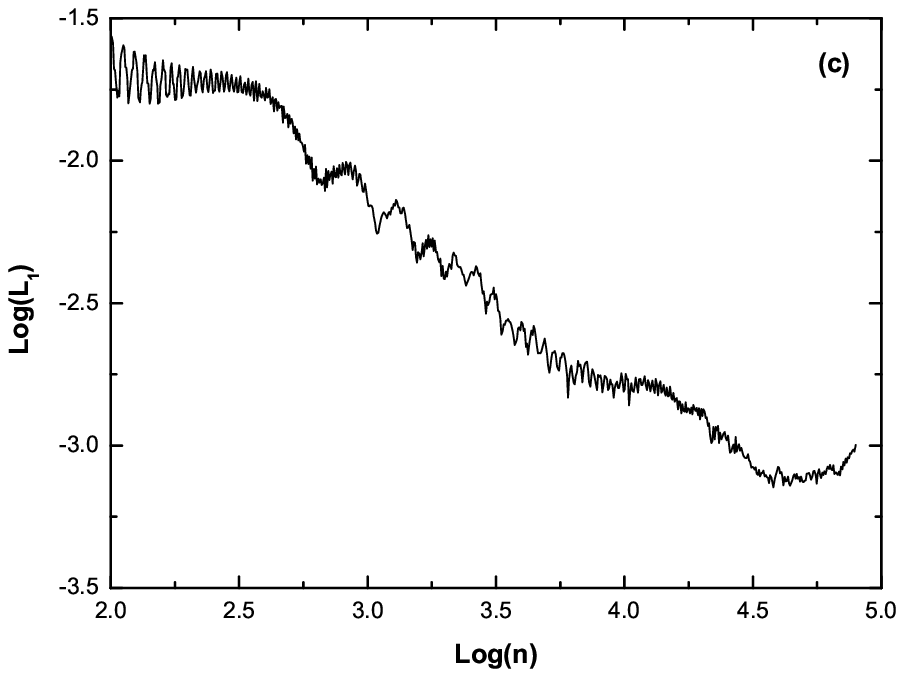} 
\end{tabular}} 
\caption{A chaotic escaping orbit of mapping (\ref{eq:map}). 
  (a) Projection of 82000  consequents of the orbit into the $x_1-x_2-x_3$ 
  space. Evolution of orbit's SALI 
  (b) and $L_1$ (c) as a function of mapping's iterations $n$ in log--log scale.} 
\label{fig:eo} 
\end{figure}

In Figure  \ref{fig:eo}(a) 
we see the projection into the $x_1-x_2-x_3$ space of a chaotic 
orbit near the edge of the beam's dynamical aperture, with initial 
condition $X^*(0)$ for $q_x=0.628615$, $q_y=0.4152$, which escapes 
to infinity after about $n=82000$ iterations. Again, the SALI 
rapidly determines the chaotic nature of the orbit as it becomes 
less than $10^{-8}$ after about $n=12000$ iterations (Figure 
\ref{fig:eo}(b)), while $L_1$ continues to decrease showing no 
sign of convergence to a non--zero value, until after $32000$ 
iterations (Figure \ref{fig:eo}(c)). 
 
This fundamentally different behavior of the SALI for ordered 
(Figure \ref{fig:oo}(b)) and chaotic orbits (Figures 
\ref{fig:co}(c) and \ref{fig:eo}(b)) and its rapid determination 
allows us to perform efficiently a more `global' study of the 
dynamics of mapping (\ref{eq:map}) in order to estimate the region 
of stability around the origin. 
 
\begin{figure} 
\includegraphics[width=7.5cm]{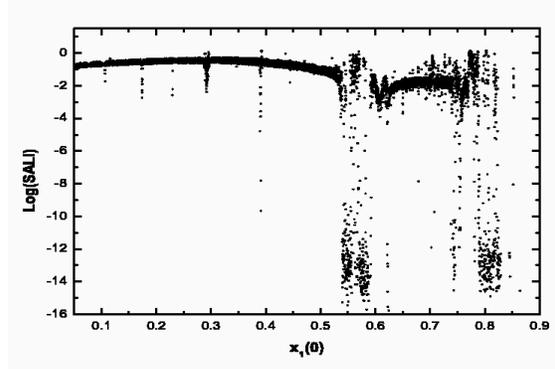} 
\caption{The values of the SALI for $n=10^5$ iterations of orbits with 
  constant initial coordinates $x_2(0)$, $x_3(0)$, $x_4(0)$ and $x_1(0) \in 
  [0, 0.9]$, as a function of $x_1(0)$.} 
\label{fig:scan_1} 
\end{figure}

As a first step in that direction, let us compute, up to $n=10^5$ 
iterations, a great number of orbits whose $x_1(0)$ coordinate 
varies from 0 to 0.9 with a step equal to $10^{-4}$, while 
$x_2(0)$, $x_3(0)$, $x_4(0)$ are the same as in the stable orbit 
of Figure \ref{fig:oo}. In Figure \ref{fig:scan_1} 
we plot the SALI of each orbit (after $n=10^5$ iterations) as a 
function of the initial coordinate $x_1(0)$. We note that chaotic 
orbits which escape in less than $n=10^5$ iterations are 
characterized as escaping orbits and are not plotted in Figure 
\ref{fig:scan_1}. From Figure \ref{fig:scan_1} we see that orbits 
with $x_1(0) \leq 0.54$ are ordered, having $\mbox{SALI} > 
10^{-4}$, except for a tiny interval around $x_1(0) \approx 0.39$ 
where one finds chaotic orbits having $\mbox{SALI} \approx 
10^{-9}$. The region $0.54 \lesssim x_1(0) \lesssim 0.59$ is 
occupied mainly by chaotic orbits having SALI values less than 
$10^{-8}$ and is followed by a region of mainly ordered motion for 
$0.59 \lesssim x_1(0) \lesssim 0.79$. For larger values of 
$x_1(0)$ chaos dominates while escaping orbits that are not 
plotted in Figure \ref{fig:scan_1} appear for $x_1(0) \gtrsim 
0.83$. 
 
Since we are interested after all in global picture of the 
dynamics around the origin, it is convenient to use the SALI 
method for `charting' this region. Let us consider therefore 
orbits with initial conditions on a grid mesh around the origin 
and evolve them for a given number $n$ of iterations. We shall 
characterize each orbit as chaotic if $\mbox{SALI}\leq 10^{-8}$ 
and as ordered if $\mbox{SALI}> 10^{-8}$. If the orbit escapes 
before the final number $n$ of iterations is reached it will be 
characterized as an escaping orbit. 
 
\begin{figure} 
\centerline{%
\begin{tabular}{c} 
\vspace{0cm} \includegraphics[width=7.5cm]{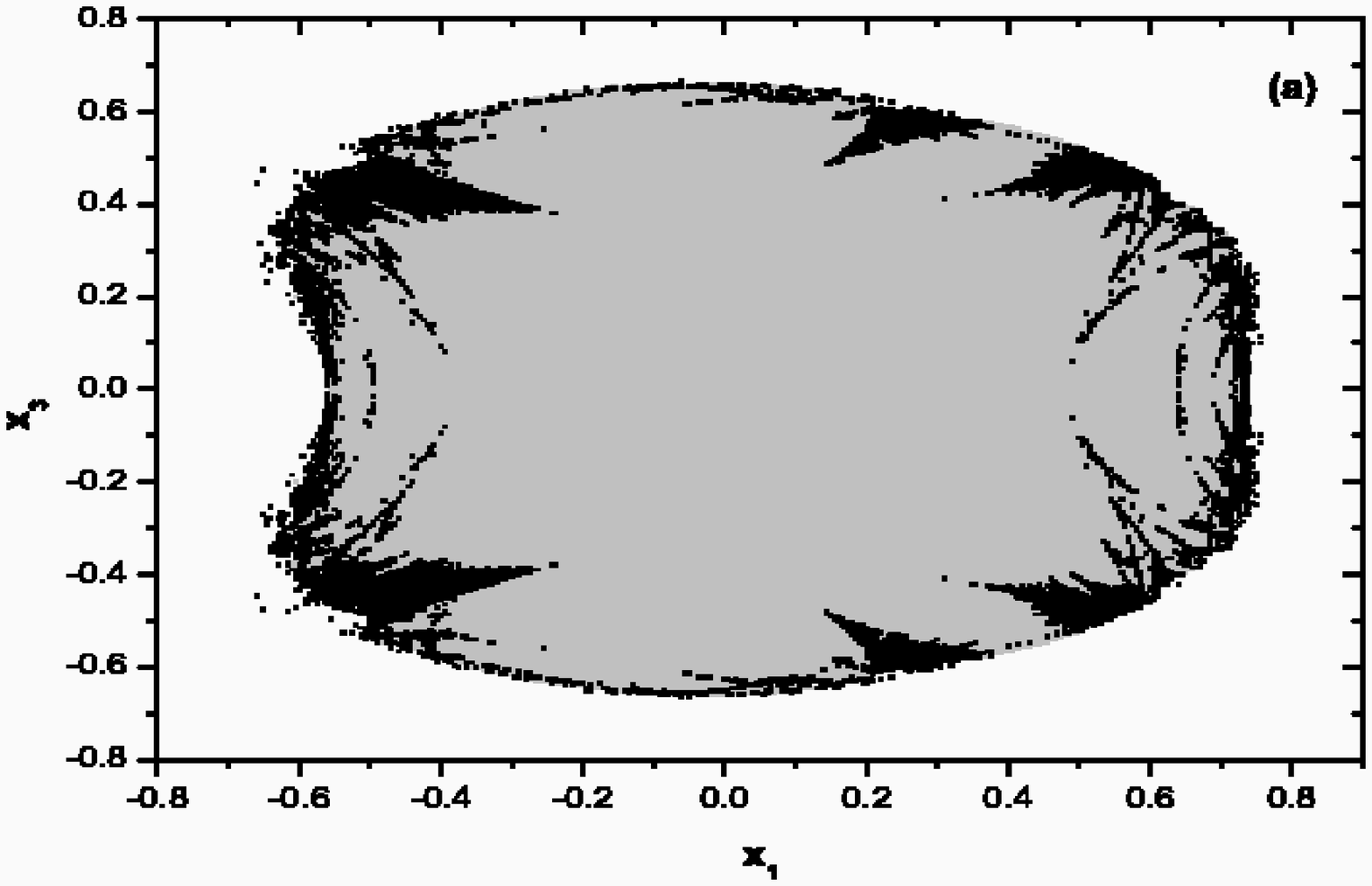} \\ 
\vspace{0cm} \includegraphics[width=7.5cm]{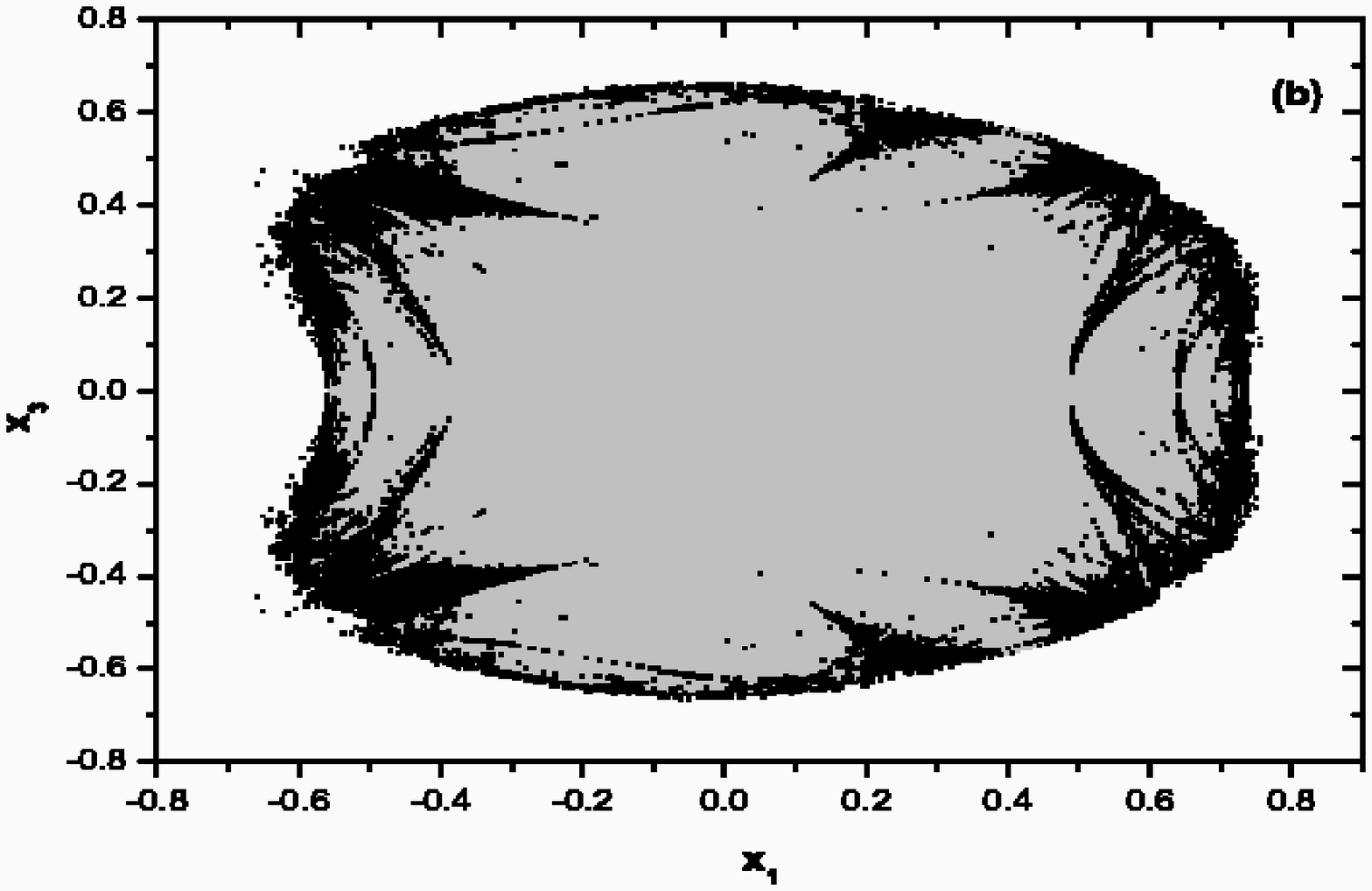} \\ 
\vspace{0cm} \includegraphics[width=7.5cm]{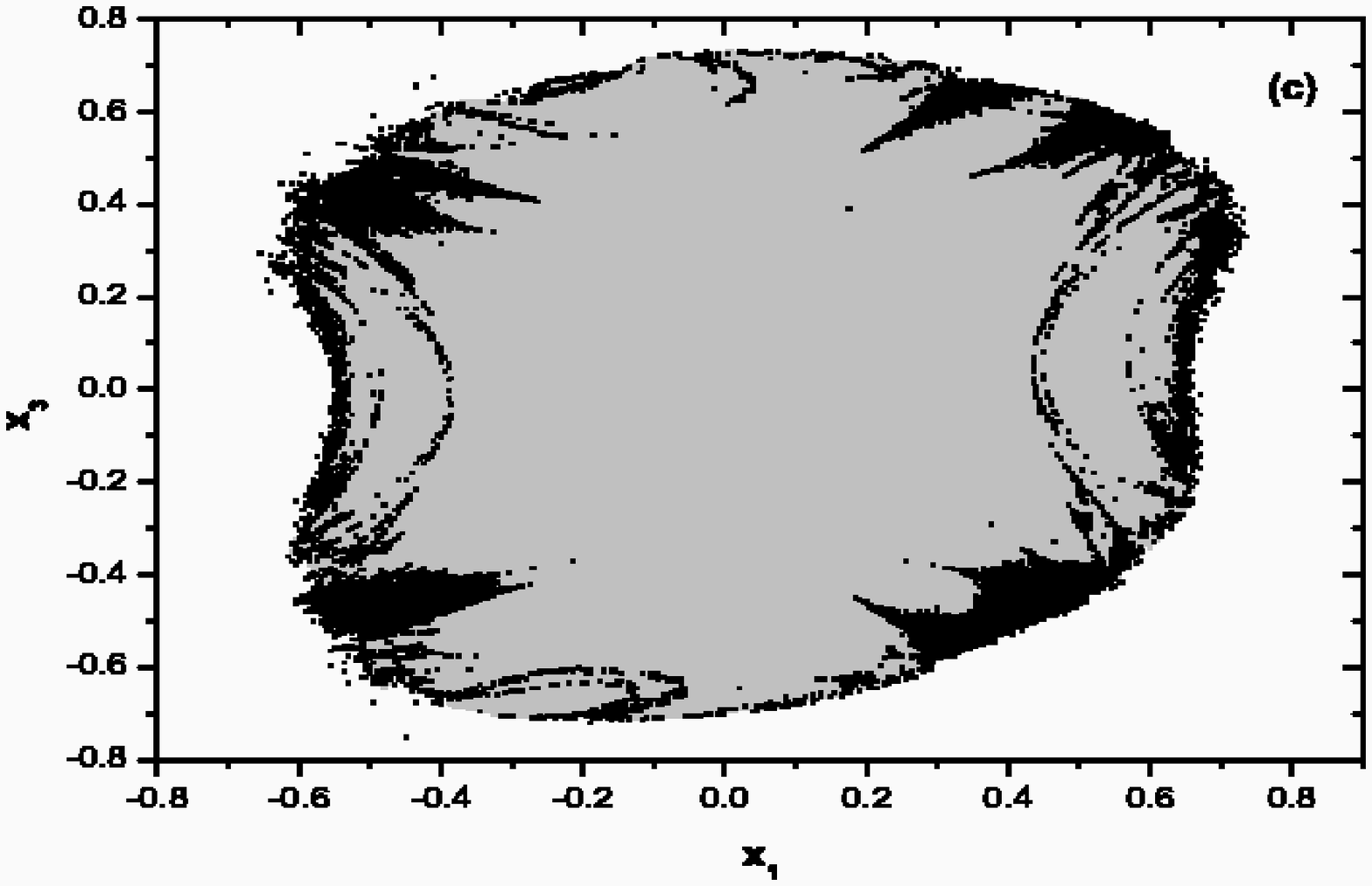} \\ 
\includegraphics[width=7.5cm]{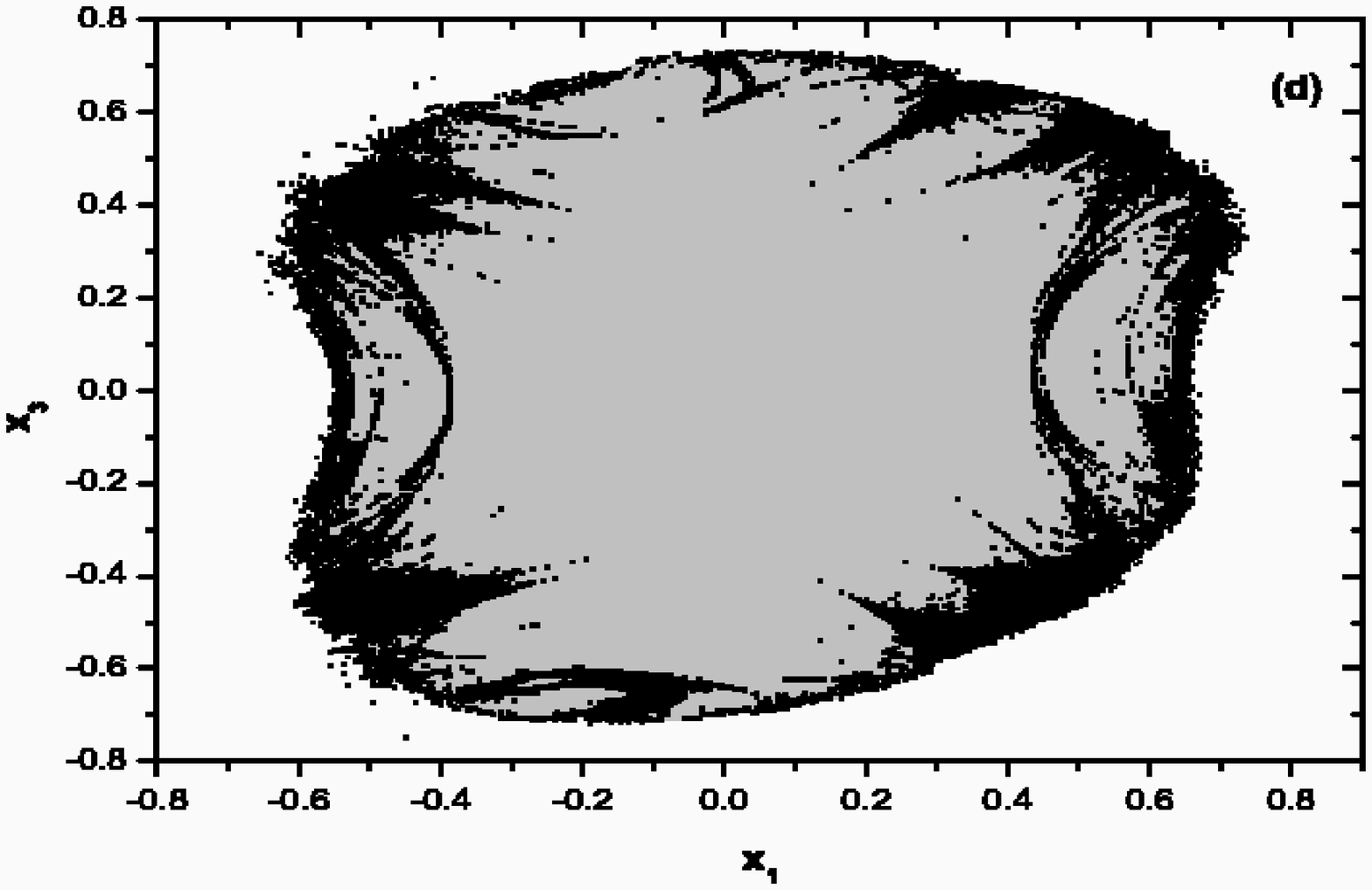} 
\end{tabular}} 
\caption{Regions of different values of the SALI on the $x_1-x_3$ 
plane after $n=10^4$ iterations (panels (a) and (c)) and after 
$n=10^5$ iterations (panels (b) and (d)). The initial conditions 
of the computed orbits on the $x_2-x_4$ plane are 
$x_2(0)=x_4(0)=0$ for panels (a) and (b) and $x_2(0)=x_4(0)=0.1$ 
for panels (c) and (d). In all frames, initial conditions are 
colored black if their $\mbox{SALI}\leq 10^{-8}$ and gray if 
$\mbox{SALI}> 10^{-8}$. The uncolored initial conditions 
correspond to orbits that escape in less than $n$ iterations.} 
\label{fig:scan_2} 
\end{figure}

We first restrict our study to the 2--dimensional configuration 
plane $x_1-x_3$ in order to be able to visualize our results. In 
particular, we consider orbits on a $400\times 400$ grid mesh 
uniformly covering the rectangular region $-1\leq x_1(0) \leq 1$, 
$-1\leq x_3(0) \leq 1$, keeping fixed the $x_2(0)$, $x_4(0)$ 
values. The corresponding `charts' are plotted in Figure 
\ref{fig:scan_2} 
for  $x_2(0)=x_4(0)=0$ (Figures \ref{fig:scan_2}(a),(b)) and for 
$x_2(0)=x_4(0)=0.1$ (Figures \ref{fig:scan_2}(c),(d)). The orbits 
were followed for $n=10^4$ (Figures \ref{fig:scan_2}(a),(c)) and 
$n=10^5$ (Figures \ref{fig:scan_2}(b),(d)) iterations in order to 
understand the evolution of `charts' with respect to the number of 
iterations. In  Figure \ref{fig:scan_2} the initial conditions 
corresponding to chaotic orbits are plotted in black while the 
ones corresponding to ordered orbits are plotted in gray. In all 
panels of Figure \ref{fig:scan_2}  all non--colored points 
correspond to escaping orbits. 
 
From the comparison of panels (a) and (b), and panels (c) and (d) 
of Figure \ref{fig:scan_2} we see that the region occupied by 
non--escaping orbits (ordered and chaotic ones) does not 
practically change as the number $n$ of iterations increases. This 
means that most of the escaping orbits fly away from the central 
region very fast, after a small number of iterations. So, the 
initial conditions plotted by black and gray color in Figure 
\ref{fig:scan_2} define the region of stability around the beam's 
circular motion in the sense that all these orbits do not escape. 
We also see that in all panels of Figure \ref{fig:scan_2} the 
region around the origin corresponds to ordered motion, while 
chaotic orbits exist mainly at the borders of the stability 
region. As the number of iterations increases, the number of 
chaotic orbits also increases. This happens because weakly chaotic 
orbits located at the borders of the region of ordered motion 
reveal their chaoticity later on as their SALI needs more 
iterations in order to become less than $10^{-8}$. Thus, although 
the number of non-escaping orbits remain practically constant the 
percentage of this number that corresponds to chaotic orbits 
increases as $n$ grows. 
 
\begin{figure} 
\includegraphics[width=7.5cm]{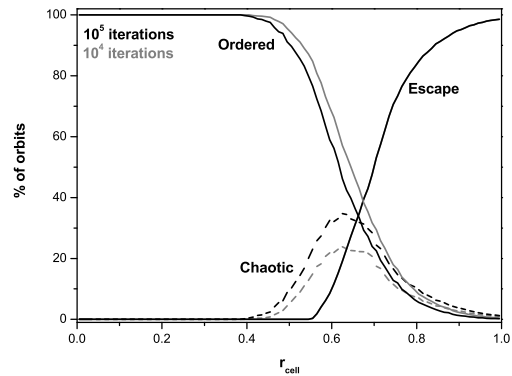} 
\caption{Percentages of ordered, escaping (solid curves) and 
chaotic (dashed curves) orbits, within spherical shells of width 
$dr=0.01$ as a function of shell's mean radius, $r_{cell}$, from 
the origin. The characterization of the orbits as chaotic or 
ordered was done according to their SALI values after $n=10^4$ 
(gray curves) and $n=10^5$ (black curves) iterations. The 
percentages of escaping orbits do not change significantly as the 
number of iterations increases and so the corresponding curves 
practically coincide.} \label{fig:scan_3} 
\end{figure}

Considering orbits with initial conditions uniformly distributed 
around the origin within a `volume' of the full 4--dimensional 
phase space, we now perform a more global analysis of orbital 
stability. As we cannot produce plots like the ones of Figure 
\ref{fig:scan_2} for the 4--dimensional space, we present in 
Figure \ref{fig:scan_3} 
the percentages of the various types of orbits within spherical 
shells of width $dr=0.01$ inside a 4--dimensional hypersphere of 
radius $r=1$ centered at the origin. We note that by the distance 
$r$ of an initial condition $(x_1(0),x_2(0),x_3(0),x_4(0))$ from 
the origin $(0,0,0,0)$ we refer to the quantity 
\begin{equation} 
r=\sqrt{x_1^2(0)+x_2^2(0)+x_3^2(0)+x_4^2(0)}\,\,. \label{eq:r} 
\end{equation} 
From the results of Figure \ref{fig:scan_3} we see again that the 
number of escaping orbits does not change significantly as the 
number $n$ of iterations increases, while the percentage of 
chaotic orbits increases with $n$. An estimation of the radius of 
the dynamical aperture therefore gives $r\approx 0.55$, as up to 
that distance from the origin no escaping orbits are found. Of 
course, for $0.4\leq r \leq 0.55$ there exists a significant 
amount of non--escaping chaotic orbits. So $r\approx 0.4$ is a 
reasonable estimate of the maximal radius of a 4--dimensional 
hypersphere around the origin, where orbits not only do not escape 
to infinity but, in addition, are also ordered. 
 
\section {Conclusions} 
\label{CONCL} 
 
In the present paper, we have applied the method of the Smaller 
Alignment Index (SALI) to the characterization of orbits of a 4D 
symplectic mapping describing the dynamics of a proton beam 
passing repeatedly through magnetic focusing elements of the FODO 
cell type. Thus, we have been able to locate efficiently islands 
of ordered motion, layers of weak chaos, as well as estimate 
accurately the beam's dynamical aperture. 
 
The success of this approach lies in the fact that it can rapidly 
distinguish between ordered and chaotic motion in Hamiltonian 
flows and symplectic mappings of any dimensionality. Since the 
SALI decays exponentially to zero in the case of chaotic orbits 
(and oscillates quasiperiodically around non-zero values in 
ordered regions), it bypasses the slow and often irregular 
convergence properties of the computation of Lyapunov exponents 
and thus provides quickly a definite characterization of each 
orbit. 
 
This allows one to use the SALI to characterize whole domains in 
phase space of different scales and sizes and `chart' them as 
chaotic or regular. Carrying out such a study for the mapping of 
this paper, we have been able to `trace out' the dynamical 
aperture of proton beams with a 2-dimensional (x and y) cross 
section, by locating 4-dimensional domains, where non-escaping 
behavior is guaranteed even after a very high number of 
iterations. Currently, we are extending our work to more realistic 
6-dimensional mappings, where longitudinal (or synchrotron) 
oscillations are taken into consideration and space charge effects 
are included \cite{BT05}. Despite the additional complications 
present in these models, we believe that the SALI method will 
again be able to yield useful results, `charting' correctly the 
dynamics of phase space domains that would otherwise be very 
difficult to probe efficiently by more traditional techniques.

\section*{Acknowledgments} 
Ch. Skokos was partially supported by the Research Committee of 
the Academy of Athens and the EMPEIRIKEION Foundation. T. Bountis 
acknowledges the partial support of the European Social Fund 
(ESF), Operational Program for Educational and Vocational Training 
II (EPEAEK II) of the Greek Ministry of Education and particularly 
the Programs "HERAKLEITOS" and "PYTHAGORAS II". 
 

 
\end{document}